\documentclass[twocolumn,prb,showpacs,preprintnumbers,amsmath,amssymb]{revtex4}
\usepackage[dvipdfm]{graphicx}
\usepackage{dcolumn}
\usepackage{bm}
\usepackage{amssymb}

\begin{document}
\title{Thermal phase transitions of supersolids in the extended Bose-Hubbard model}

\author{Kwai-Kong Ng}
\affiliation{Department of Physics, Tunghai University, Taichung, Taiwan}
\date{\today}
\begin{abstract}
We investigate numerically the finite-temperature phase diagrams of the extended Bose-Hubbard model in a 
two-dimensional square lattice. In particular, we focus on the melting of supersolid phases of two different crystal orderings, stripe and star orders, arising from the competition of the nearest- and next-nearest -neighbor interactions in the vicinity of quarter filling.
The two crystal orders are the result of broken translational symmetry in either one or in both $x$, and $y$ directions.
The broken gauge symmetry of the supersolids are found to be restored via a Kosterlitz-Thouless transition while the broken translational symmetries are restored via a single second-order phase transition, instead of two second-order transitions in the Ising universality class. On the other hand, the phase transitions between the star and stripe orders are first order in nature.

 \end{abstract}


\maketitle
\section{Introduction}
Supersolids with both diagonal and off-diagonal long-range order are observed in various model, 
either with or without hard-core constraint.\cite{Batrouni,Sengupta,Wessel1,Chen,Ng1,Dang,Ng2} One obvious necessary condition for the occurrence of 
supersolids (SSs)
is the presence of a small but finite kinetic energy in competition with a relatively large repulsive 
interactions. Previous results suggest that the hard-core constraint may destroy supersolid phases by 
the formation of domain walls, which leads to phase separation instead.\cite{Batrouni} Stable supersolid phases in hard-core 
models are found, however, in systems contain frustrated interactions.\cite{Batrouni,Wessel1,Chen,Ng1,Dang,Ng2}

The extended Bose-Hubbard model is 
a typical example that has been proposed to demonstrate a supersold phase within the mean-field approximation \cite{Otterlo}, 
and is confirmed by quantum Monte Carlo simulations.\cite{Batrouni,Chen,Ng1,Dang,Ng2,Schmid} As long as the next-nearest-neighbor (nnn) 
interaction $V_2$ is dominant, hard-core bosons lining up as stripes to reduce potential energy near 
half filling \cite{Batrouni} (see Fig. \ref{ordering}).
At quarter filling, frustrations induced by competing nearest neighbor (nn) $V_1$ can lead to a stable star
 order phase that characterized by 
 finite 
structure factors at wave vector $(\pi,\pi)$ and $(\pi,0)$ [$(0,\pi)$]. The existence of the 
corresponding supersolid is also numerically confirmed recently.\cite{Chen,Ng1,Dang,Ng2}

The melting of stripe supersolids has been
investigated in Ref. 9 where the finite-temperature phase diagrams are determined. At half filling, 
the stripes melt at a first order phase transition, while in the doped system, the melting transitions of the solid and supersolid are either very weakly first order or of second order. Both the broken rotational and translational symmetries of the stripe phases are found to be restored at a single continuous transition and no intermediate nematic (liquid crystal) phase has been observed. Nevertheless, Ref. 9 focuses solely on 
the case with no nn interaction $V_1$ and so only the melting of stripe phases are discussed. 
The thermal melting behavior of the star supersolid phases near the quarter filling is then largely unknown.
Since the star order is characterized by two structure factors, the two broken translation symmetries can, in principle, be restored simultaneously or at two different temperatures. The latter scenario means the star phase melts into a stripe phase before changes to a normal fluid or superfluid. It is also worth to examine whether the missing nematic phase in Ref. 9 can be stabilized by the introducing $V_1$ in the system.  
It is the purpose of this work to 
address these issues and determine the finite-temperature phase diagram for the competing nn and nnn interactions.

\section{Model}

To start with, the extended Bose-Hubbard model on a two-dimensional square lattice with the
Hamiltonian is defined as
\begin{equation}
H_b =  -t\sum_{i,j}^{nn} (b^\dagger_i b_j+b_i b^\dagger_j) + V_1\sum_{i,j}^{nn} n_i n_j + V_2\sum_{i,j} ^{nnn}n_i n_j - \mu \sum_{i} n_i
\end{equation}

\noindent with hard-core constraint $n_i\leq1$. In the above formula, $b (b^\dagger)$ is the boson destruction (creation) operator
and $\sum^{nn} (\sum^{nnn})$ sums over the nearest (next-nearest) neighboring sites. The energy scale is fixed by
$t=1$ throughout this paper.

\begin{figure}
\includegraphics[width=7.5cm]{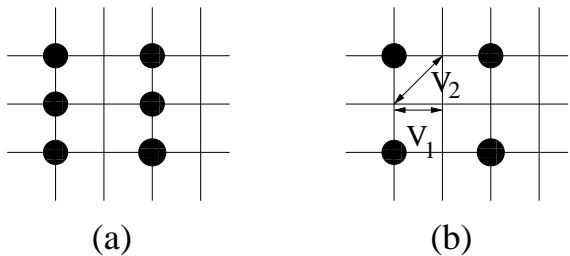}
\caption{A schematic description of the classical (a) stripe order and (b) star order for half and quarter fillings, respectively. While there is only one broken translational symmetry in the stripe phase, two translational symmetries in both directions are broken in the star phase. On the other hand, the $x$-$y$ rotational symmetry is broken in the stripe phase but not in the star phase.}
\label{ordering}
\end{figure}

In the case of $V_2=0$, it is well known that the ground state at half filling is a checkerbroad state, which becomes a superfluid as the system is doped away from half filling. On the other hand, in the case
of $V_1=0$, the half-filled ground state has a stripe order that characterized by the finite structure factors
at wave vector {\bf Q}$=(\pi,0)$ or {\bf Q}$=(0,\pi)$ (Fig. \ref{ordering}). More interesting is the coexistence of the stripe order 
and superfluidity, i.e., the stripe supersolid phase, in the doped system. Most condensed bosons in this 
exotic states move in the direction parallel to the stripes but superflow in the perpendicular direction is
also observed.\cite{Batrouni} The stripe quantum solid proceeds to the superfluid via two second-order transitions, in contrast to the case of vanishing $V_2$, where transitions from quantum solid to superfluid are discontinuous.  

The phase diagram becomes even richer when both $V_1$ and $V_2$ are finite. An extra quarter-filled solid (see Fig. \ref{ordering}) and its corresponding supersolid are predicted within 
the mean-field theory \cite{Otterlo} and are confirmed by numerically simulations. \cite{Batrouni,Chen,Ng1,Dang,Ng2,Schmid} Quarter-filled solids with two distinct 
crystal structures are possible depending on the competition of $V_1$ and $V_2$, while only one
structure is stable for the supersolid phase.\cite{Ng2} This so-called star supersolid has been discussed in
detail but the finite-temperature phase diagram has not been examined so far and therefore serves 
as the subject of this work.

One of the most efficient approach to study the bosonic models is the standard stochastic series expansion (SSE) Monte Carlo method implemented with a directed loop algorithm.\cite{Sandvik} In SSE, the superfluidity, given by $\rho_{x(y)}=\langle W_{x(y)}^2\rangle/4\beta t$, is computed by measuring the winding number fluctuation within the simulations. Hereafter we present the average superfluidity $\rho_s=(\rho_x+\rho_y)/2$. The broken translational symmetries in the $x$ and $y$ directions
of the striped phases are 
characterized by the structure factors  

\begin{equation}
S(\textbf{Q})=\sum_{ij}\langle n_i n_j e^{i \textbf{Q}\textbf{r}_{ij}}\rangle/N^2,
\end{equation}

\noindent at the wave vectors $\textbf{Q}=(\pi,0)$ and $(0,\pi)$, respectively. For convenience, we measure
the average value of 

\begin{equation}
S_+=[S(\pi,0)+S(0,\pi)]/2, 
\end{equation}

\noindent in order to avoid the difficulty to determine the direction of broken
translational symmetry in each simulation. Moreover, we also study the rotational symmetry breaking
in the local-density correlation by measuring the order parameter $O_N$,

\begin{equation}
 O_N=\sum_{x,y} n_{x,y} (n_{x+1,y}-n_{x,y+1}),
\end{equation}

\noindent with $n_{x,y}$ the boson number operator at the site $(x,y)$. We stress that $O_N$ signals the local broken rotational symmetry but not the global one.
The star order phase, on the other hand, is characterized by the finite 
values of both $S(\pi,\pi)$ and $S_+$, which breaks the translational symmetry in both directions.

\section{Ground-state phase diagram}

\begin{figure}
\includegraphics[width=7.5cm]{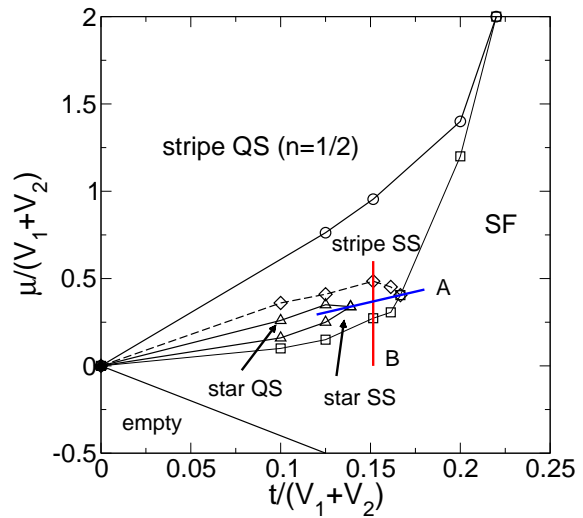}
\caption{(Color online) Ground-state phase diagram for $V_1=2.0$. It contains a half-filled stripe quantum solid and a quarter-filled star quantum solid and their corresponding SSs. Finite-temperature 
phase diagrams scan along the lines A and B are presented in Figs. \ref{QF} and \ref{FTP}, respectively. The dotted line indicates a first-order transition from the star supersolid to stripe supersolid. The system size used here is $L=28$ with $T=1/2L$.}
\label{GS}
\end{figure}

The ground-state phase diagrams $V_1$ vs $V_2$ at quarter filling have been presented before \cite{Ng1,Ng2}. The star 
supersolid is found to be stable for a wide region of parameters $V_1$ and $V_2$ even at commensurate filling
$n=1/4$. For the following discussion, we choose a representative nn interaction $V_1=2.0$ of which the 
phase diagram $\mu$ vs. $t$ is shown in Fig. \ref{GS}. Because of the particle-hole symmetry, the 
symmetric upper half of the diagram is not shown. The main lobe in the diagram is the striped 
quantum solid phase at half filling and from which decreasing the boson number leads to a striped supersolid phase. 

As mentioned before, the striped supersolid is characterized by a finite $S_+$ that 
breaks one translational symmetry but preserves the translational symmetry in another direction (Fig. \ref{ordering}). 
Around quarter filling, the $S(\pi,\pi)$ becomes  finite as the system enters the star supersolid phase. The 
translational symmetry is now broken in both $x$ and $y$ directions.
At exactly quarter filling the superfluidity $\rho_s$ vanishes for $V_2 \gtrsim 4.9$ (see Fig. \ref{QF}) but remains finite otherwise.

As shown in the Fig. 1 of Ref. 8, the general features of the ground-state phase diagram can also be described by the mean-field theory.  The quarter-filled star quantum solid phase is enclosed by a narrow region of star supersolid phase before enters the stripe supersolid phase. However, the mean-field approximation tends
to overestimate the extent of the stripe and star supersolid phases as presented in the Fig. 8 of Ref. 7. This indicates that the neglected quantum fluctuations in the mean-field calculations destabilize the 
supersolid phases and reduce the region of the supersolid phases in the ground-state phase diagram.

In the following, we present the finite-temperature phase diagrams at quarter filling and away from quarter 
filling along the lines A and B in Fig. \ref{GS}, respectively.

\section{Finite-temperature phase diagram at quarter filling}

\begin{figure}
\includegraphics[width=7.5cm]{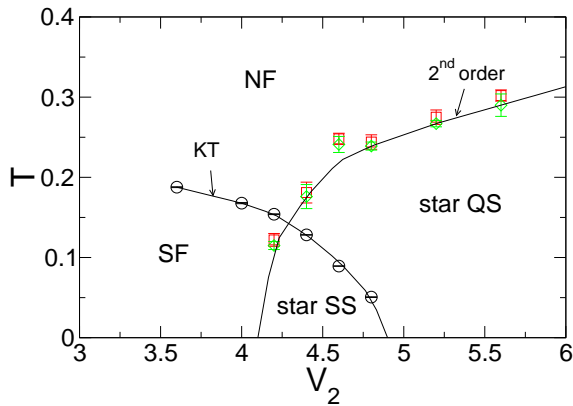}
\caption{(Color online) Finite-temperature phase diagram at quarter filling along the line A in Fig. \ref{GS}. The black solid lines are guides to the eyes. Data points are obtained from the finite-size analysis as described in Figs. \ref{FSa} and \ref{BC1}.  In the following figures, error bars are smaller the symbol size unless otherwise stated.}
\label{QF}
\end{figure}

Figure \ref{QF} shows the finite-temperature phase diagram with fixed boson density of 1/4 (line A
in Fig. \ref{GS}). In our grand-canonical simulation we adjust the chemical potential $\mu$ in order to obtain the average particle
density $n$ equals to quarter filling. At zero temperature, the off-diagonal long-range order is depressed 
when $V_2$ increases,
while the diagonal long-range order, the star order in this case, is enhanced. In the intermediate $V_2$, both orderings coexist and gives rise to a star supersolid phase. The presence of the supersolid phase is in contrast to $n$=1/2 case where the superfluid proceeds to the stripe quantum solid via a first-order transition with no intermediate supersolid phase. In Fig. \ref{QF}, the superfluid changes to the star quantum solid via two second-order transitions instead. For the study of thermal transitions, we carry out a finite-size analysis on the order parameters $\rho_s$, $S_+$, and $S(\pi,\pi)$. 

\begin{figure}
\includegraphics[width=7.5cm]{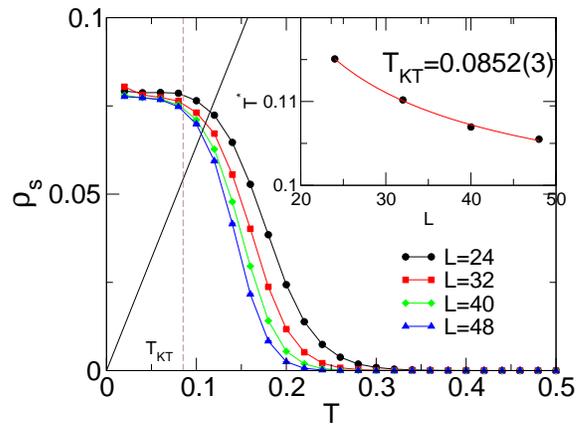}
\caption{(Color online) Finite-size scaling of the superfluidity $\rho_s$ for $\mu/(V_1+V_2)=0.394$. The inset shows the logarithmic correction $T^*=T_{KT}\{1+1/[2ln(L/L_0)]\}$.}
\label{FSa}
\end{figure}

In a two-dimensional system, the superfluid to normal-fluid
phase transition can be well described by the Kosterlitz-Thouless (KT) transition.\cite{Kosterlitz} Plotting the superfluidity $\rho_s$ as a function of $T$, as shown in Fig. \ref{FSa} for a representative $V_2$, the intersection 
values of 
$\rho_s(T^*)=\frac{2}{\pi}T^*$ for different sizes $L$ follow the logarithmic correction $T^*=T_{KT}\{1+1/[2ln(L/L_0)]\}$. Fitting to this relation gives the value of critical temperature $T_{KT}$ for infinite system.
Indeed, the values of $T^*$ shown in the inset of Fig. \ref{FSa} follow very well the logarithmic correction and confirms the superfluid to normal fluid transition is of the KT type. The critical temperature T$_{KT}$, as shown in Fig. \ref{QF}, reduces smoothly as $V_2$ increases and disappears for $V_2\gtrsim4.9$. 

\begin{figure}
\includegraphics[width=7.5cm]{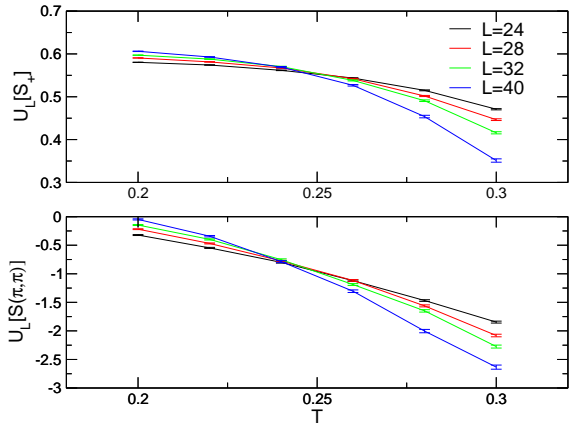}
\caption{(Color online) Binder cumulant of $S_+$ and $S(\pi,\pi)$ as a function of temperature for different system sizes ($V_1=2.0$, $V_2=4.6$).}
\label{BC1}
\end{figure}

Next, we analyze the temperature dependence of the structure factors $S_+$ and $S(\pi,\pi)$. As mentioned before, the two broken translational
symmetries can be restored at two different temperatures or restored at the same temperature. To
determine the transition temperatures, we recall that for second-order transitions, the fourth order Binder cumulant ratios $U_L=1-\langle O^4 \rangle /3 \langle O^2 \rangle ^2$ should intersect at the same critical point for different sizes $L$.\cite{Binder} Figure \ref{BC1} shows the cumulant ratios of $S_+$ and $S(\pi,\pi)$, respectively. Indeed, $U_L$ meet at the same transition points independent of the system size. 
Particularly, the resulted transition points from both cumulant ratios are very close to each other and are indistinguishable within the statistical uncertainties of our data. Our data thus suggest a single second-order
transition for the melting of the star quantum solid to the normal-fluid phase, or a very narrow intermediate stripe quantum solid in between
the phases.

\begin{figure}
\includegraphics[width=7.5cm]{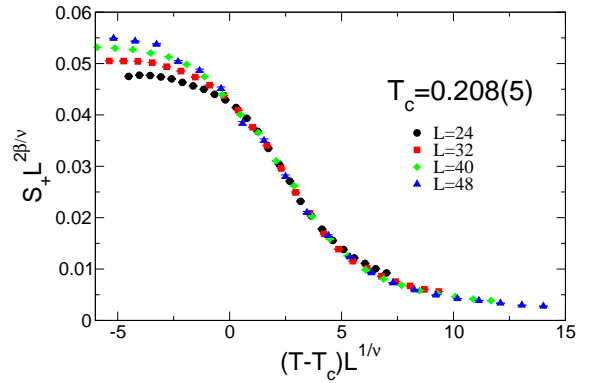}
\caption{(Color online) Attempt of the finite-size scaling for the structure factor $S_+$ ($\mu/(V_1+V_2)=0.394$) by assuming the critical exponents $\beta=1/8$ and $\nu=1$ of the Ising universality class. At low temperatures, significant derivations from the scaling law indicate the transition is not of the Ising type.}
\label{FS1}
\end{figure}

\begin{figure}
\includegraphics[width=7.5cm]{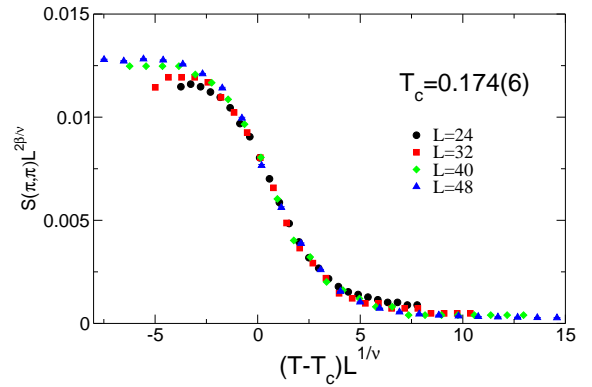}
\caption{(Color online) Finite-size analysis of $S(\pi,\pi)$.}
\label{FS2}
\end{figure}

To further clarify these two scenarios, we suppose there are two second-order phase transitions that restores the translation symmetries sequentially. We expect both transitions are in the three-dimensions Ising universality class as each transition breaks a Z$_2$ symmetry. Therefore the structure factors should follow the finite size scaling law $S=L^{-2\beta/\nu}F[(T-T_c)L^{1/\nu}]$ with $\beta=1/8$
and $\nu=1$ as expected for the Ising class.\cite{Fisher,Laflorencie} The data of $SL^{2\beta/\nu}$ for different $L$ should intersect at the same temperature that determines the value of
$T_c$. By rescaling the temperature to $(T-T_c)L^{1/\nu}$, all data will collapse into a single line, 
provided the data are close enough to the critical point where the scaling law is valid. As shown in Figs. \ref{FS1} and \ref{FS2}, the data of $S_+$ and $S(\pi,\pi)$ intersect at a size-independent critical temperature, respectively. The critical temperature of $S(\pi,\pi)$ is found significantly lower than that of the $S_+$
and therefore suggests an intermediate stripe quantum solid phase. However, the rescaled data collapse well only for $T>T_c$ but a clear derivation from the scaling law is observed for temperatures below $T_c$. This failure 
of data collapse invalidates the assumption of two transitions of Ising type. Therefore it rules out the possibility of a very narrow intermediate stripe quantum solid phase lies between the star quantum solid and normal-fluid phases. The melting of the star quantum solid to normal fluid is then likely to be a single second-order transition. The system happens not to choose to restore the translational symmetry in one particular direction first and the other direction later, but it restores both broken translatonal symmetries  simultaneously. We will comment on this in the next section.

\section{Finite-temperature phase diagram away from quarter filling}

\begin{figure}
\includegraphics[width=7cm]{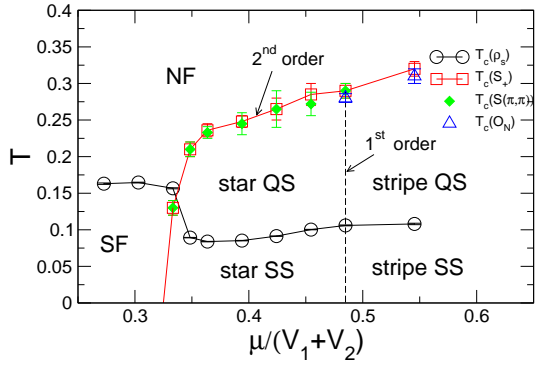}
\caption{(Color online) Finite-temperature phase diagram ($V_1=2.0$, $V_2=4.6$) as a function of $\mu$ along the line B in Fig. \ref{GS}.}
\label{FTP}
\end{figure}

\begin{figure}
\includegraphics[width=7.5cm]{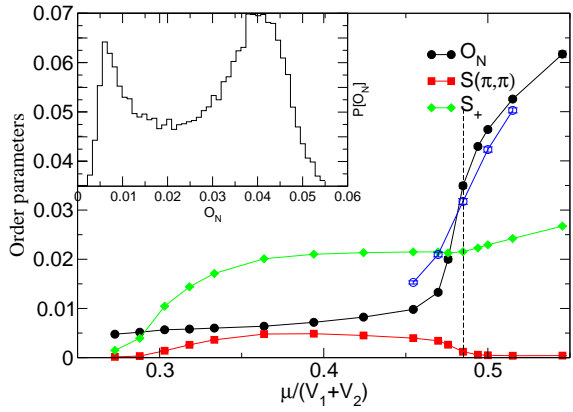}
\caption{(Color online) Order parameters as a function of $\mu$ ($V_1=2.0$, $V_2=4.6$, $L=40$, $T=0.02$). The parameters $O_N$ are plotted for sizes $L=40$ (filled circles) and $L=28$ (open circles) to demonstrate the finite-size effect. Inset: histogram of the parameter $O_N$ at $\mu/(V_1+V_2)=0.4848$ shows a double peak feature of the first-order transition between the star and the stripe supersolid phases.}
\label{hist}
\end{figure}

When the system is doped away from the quarter filling, the finite-temperature phase diagram is even richer. Figure \ref{FTP} presents the finite-temperature phase diagram, determined by the Binder cumulants as 
discussed previously, as a function of chemical potential $\mu$ with fixed $V_1=2$ (line B in Fig. \ref{GS}). 
The zero-temperature ground states are the superfluid, star supersolid, and stripe supersolid phases, respectively, as the boson number is increased. The quantum phase transition of the superfluid to star supersolid is second order, whereas the transition from the star supersolid to the stripe supersolid is first order. Figure \ref{hist} plots the ground-state order parameters as a function of $\mu$ from which an abrupt change is observed at the transition between two supersolids. The histogram of $O_N$ (see the inset) shows the double peaks feature as 
a clear signature for a first-order transition. This discontinuous transition reflects 
the fact that the star order is more than simply breaking another translational symmetry of the stripe order.
It implies the two broken translational symmetries in the star order are not independent and 
therefore cannot be constructed by breaking the translational symmetries
sequentially via two second-order transitions.  This also explains the direct melting of the star quantum solid observed in Fig. \ref{QF}.

As the temperature is increased, again, the broken gauge symmetry is restored via a KT transition.
However, in contrast to the Fig. \ref{QF} where the critical temperature $T_c(\rho_s)$ reduced smoothly without being affected by the emergence of broken translational symmetries,  $T_c(\rho_s)$ in Fig. \ref{FTP} exhibits a sudden drop in the star order phase. The $T_c(\rho_s)$ remains rather flat throughout the whole star supersolid phase. This sudden suppress of superfluidity is thus unlikely due to the increase in particle number alone, but may be resulted by the coupling of the broken translational symmetries. This is also in contrast to the supersolid melting in most systems \cite{Laflorencie,Schmidt,Yamamoto,Boninsegni} where the restoration of superfluid symmetry and translational symmetry are largely independent. This interesting behavior may be related to the intrinsic structure of the star supersolid and further study is needed to gain more insight of this particular phase.

\begin{figure}
\includegraphics[width=7.5cm]{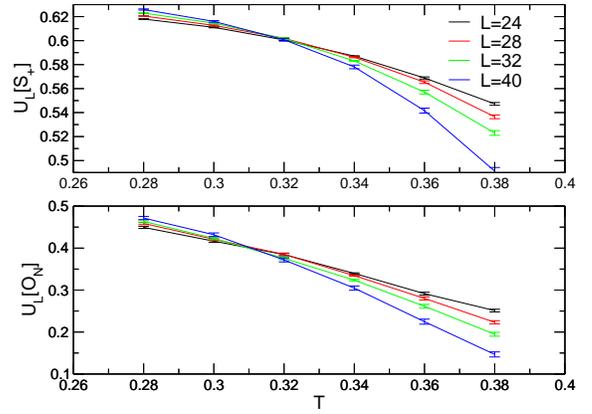}
\caption{(Color online) Binder cumulant of $S_+$ and $O_N$ as a function of temperature for different system sizes ($V_1=2.0$, $V_2=4.6$).}
\label{BC2}
\end{figure}

For larger chemical potential, and particle density as well, we found that the stripe supersolid melts into a stripe quantum solid and then further into the normal fluid through a second-order transition. From the plots of Binder cumulants for both $S_+$ and $O_N$ in Fig. \ref{BC2}, we observe that the transition temperatures for both broken translation and rotational symmetries  are within our data uncertainties and again suggests a single second phase transition. Therefore no intermediate  nematic phase is observed. While this finding is consistent to the case of vanishing nn interaction $V_1$, whether larger $V_1$ will be able to stabilize the nematic phase is worth to study.

\section{Summary}
We present the finite-temperature phase diagrams of the extended Bose-Hubbard model in the vicinity of the quarter-filling. The transitions from the super to normal phases that restore the gauge symmetry are of the KT 
type in nature. Interestingly, the transition temperature exhibits a sudden drop when the star order emerges. It differs from the usual supersolid that the superfluid $T_c$ behaves
 largely independent of the crystal order of the system. Futher exploration of the possible coupling between
the broken gauge and translational symmetries should provide more insight of this behavior.

For the melting of star order, our result support the scenario of a single second-order transition instead of
 two Ising-type second-order transitions. Translational symmetries are restored simultaneously and no
 intermediate phase of stripe order is found. Our result indicates that the two order parameters of the star
 order couple together and the state cannot be viewed simply as two interwined stripe orders. 
It will be interesting to investigate the effect of a small anisotropy in one dimension that lifts
the degeneracy of the $x$, and $y$ symmetry. As one translational symmetry is more robust, the critical temperatures of the broken translational symmetries should be different and the melting of star quantum solid
might occur as two second-order transitions. The effect of anisotropy on the stripe order melting will also be worth to study to see if the nematic state can be stabilized near the stripe order.

\begin{acknowledgments}
We thank Y. C. Chen and M. F. Yang for stimulating discussions. The numerical computations are performed in the Center for High Performance Computing of the THU.  K.K.N. acknowledges the support by the National Center for the theoretical Science and the finanical support from the NSC (R.O.C.), Grant No. NSC 97-2112-M-029-003-MY3.
\end{acknowledgments}

\end{document}